\documentclass[12pt,a4page]{iopart}
\usepackage{textcomp}
\usepackage{graphicx}
\begin{document}

\title[Galileon in Collisions]{Horndeski/Galileon in High Energy Collisions}

\author{B. N. Latosh${}^{1,2,3}$}
\ead{latosh.boris@gmail.com}
\address{${}^1$ Bogoliubov Laboratory of Theoretical Physics, Joint Institute for Nuclear Research, 141980 Dubna, Joliot-Curie 6, Moscow region, Russia}
\address{${}^2$ State University ''Dubna'', 141980 Dubna, Universitetskaya st. 19, Moscow region, Russia}
\address{${}^3$ Sternberg Astronomical Institute, Lomonosov Moscow State University, Universitetsky Prospekt, 13, Moscow 119991, Russia}

\begin{abstract}
Horndeski/Galileons may be considered as a proper generalization of General Relativity in high energy regime. Thus one may search for manifestation of Galileons interaction in collision experiments. In this paper we give arguments supporting this thesis. Galileon scalar field do not interact with matter via Standard Model interactions, we discuss a mechanism that allows Galileons to have influence on particle collisions. We give reasons to narrow the whole class of Horndeski/Galileons models to one particular term -- John term from Fab Four subclass -- for this particular issue. We were able to establish the constraint on the model coupling constant $\beta < 10^{36}$ GeV${}^{-2} =  10^{-13} $ g${}^{-2}$.
\end{abstract}




\section{Introduction}

Horndeski/Galileons are the most general class of scalar field models with second-order field equations. They were discovered by Gregory Horndeski in 1974 \cite{Horndeski:1974wa} and later were formulated in terms of Galileons interactions \cite{Kobayashi:2011nu} (see \cite{Deffayet:2013lga} for a review). Galileons are considered as scalar-tensor modification of General Relativity (GR) thus it is important to establish if Galileons are able to describe any beyond GR effects.

Galileons possess a few attractive features. First, as Galileons have second-order field equations they are free from Ostrogradski instability. It is worth mentioning that Galileons are not the most general class of scalar-tensor models free from the instability. Such models are known as ''Beyond Horndeski'' and were discovered in \cite{Zumalacarregui:2013pma,Gleyzes:2014dya}. Beyond Horndeski models have high order field equations, though true propagating degrees of freedom obey second-order equations, thus avoiding Ostrogradski instability. Second, Galileons admit screening mechanisms \cite{Kase:2013uja,Nicolis:2008in,Babichev:2013usa} thus mimic GR on the scale of star systems. Third, Galileons may drive inflation via non-trivial kinetic coupling \cite{Kobayashi:2010cm,Kobayashi:2011nu,ArmendarizPicon:1999rj}. Finally, there exists a special subclass of Galileons known as ''Fab Four'' \cite{Charmousis:2011bf,Charmousis:2011ea}. Fab Four models admit self-tuning solutions and they are able to completely screen the curvature of spacetime from the net cosmological constant. The models can violate Weinberg no-go theorem \cite{Weinberg:1988cp} via violation of Poincare invariance on the self-tuning vacuum, while they are still able to sustain flat spacetime. Finally, recently it was shown that some models from Fab Four class are capable of describing both inflation and late-time acceleration of the Universe \cite{Starobinsky:2016kua}.

All this features provide evidences that Galileons (and especially Fab Four) may be used as proper generalization of GR in high energy regime. At the same time Galileons are non-renormalizable \cite{Luty:2003vm,Hinterbichler:2010xn} and may not be treated as a theory of quantum gravity. Thus we will consider Galileons as effective theory that acts on the energies between GR and Quantum Gravity. This allows one to search for manifestation of Galileons in order to constraint model parameters.

First approach to this problem is based on low energy manifestation of Galileons field. One may study black hole-like solutions \cite{Rinaldi:2012vy,Babichev:2013cya,Charmousis:2015aya,Babichev:2015rva,Anabalon:2013oea,Minamitsuji:2013ura,Babichev:2016fbg} (see \cite{Babichev:2016rlq} for a more detail review) like it was done in \cite{Tretyakova:2016knb}. This study may result in strong constraints similar to the results of the papers \cite{Tretyakova:2016knb,Tretyakova:2015vaa}. For Galileons this may not be the case, as Fab Four admits so-called stealth Schwarzschild solution \cite{Babichev:2013cya}. The solution has Schwarzschild metric and non-trivial configuration of Galileon field thus it may not be distinguished from standard Schwarzschild solution by geodesic motion (This issue was also pointed in \cite{Tretyakova:2016knb}). We also would like to highlight that Horndeski/Galileons models were widely studed within classical physics framework. Stability of Galileons black holes was also considered in many papers like \cite{Minamitsuji:2014hha,Cisterna:2015uya}. And constraint on time variation of effective gravitational constant was obtained in a paper \cite{Kimura:2011dc}.

Alternative approach lies in the area of high-energy physics, one may search for signatures of Galileons in particle collision experiments. As we treat Galileons as leading correction to GR in high energy regime one may expect to find such signatures in LHC data. This approach has two issues. First, Galileons are non-renormalizable theory \cite{Luty:2003vm,Hinterbichler:2010xn} thus one is limited by tree-level diagrams. Second, Galileons do not interact with usual matter via Standard Models interactions, thus Galileons may manifest themselves only in gravitational processes. So, all Galileons-involving processes would be strongly suppressed. On the contrary, nontrivial Galileons-graviton interaction will influence standard scalar-graviton vertex and may compensate the smallness of gravitational coupling. One only needs to find a Fab Four model which would be able to provide such a compensation.

\section{Galileons production in particle collisions}

The Fab Four models are formed by the following terms:
\begin{eqnarray}\label{F_Four}
  L_{John} & = & \sqrt{-g} ~ V_{John} (\phi) ~ G^{\mu\nu} \nabla_\mu \phi  \nabla_\nu \phi  \\
  L_{Paul} & = & \sqrt{-g} ~ V_{Paul} (\phi) ~ P^{\mu\nu\alpha\beta} \nabla_\mu \phi \nabla_\nu \phi \nabla_\alpha \nabla_\beta \phi \\
  L_{George} & = & \sqrt{-g} ~ V_{George}(\phi) ~R  \\
  L_{Ringo} & = & \sqrt{-g} ~ V_{Ringo}(\phi) ~ \hat G \label{F_Four_end}
\end{eqnarray}
where $G_{\mu\nu} = R_{\mu\nu}-\frac{1}{2} R g_{\mu\nu}$ is the Einstein tensor; $\hat G = R_{\mu\nu\alpha\beta} R^{\mu\nu\alpha\beta} - 4 R_{\mu\nu} R^{\mu\nu} + R^2$ is the Gauss-Bonnet term; $P^{\mu\nu\alpha\beta} = -\frac14 \varepsilon^{\mu\nu\lambda\sigma} \varepsilon^{\alpha\beta\rho\gamma} R_{\lambda\sigma\rho\gamma}$ is the double dual Riemann tensor; $\varepsilon^{\mu\nu\alpha\beta}$ is the Levi-Civita tensor and $\phi$ is the Galileon field. Terms $V_{John}$, $V_{Paul}$, $V_{George}$, $V_{Ringo}$ are arbitrary potentials depending only on Galileon field $\phi$. 

Lagrangians \eref{F_Four}-\eref{F_Four_end} provide an infinite number of possible models, thus we need to define principles that would allow us to narrow this number. First of all, we would like to consider only the model with non-trivial kinetic coupling. A model of that class may violate null energy condition providing some analogy of inflation \cite{Nicolis:2009qm,Creminelli:2010ba,Creminelli:2012my}. George and Ringo terms have no non-trivial kinetic coupling so we will exclude them. One should also remember that George term is equivalent to GR at the level of the action and Ringo term belongs to Einstein-dilaton-Gauss-Bonnet class of scalar-tensor models which are the topic of different discussion lying beyond the scope of this paper. Second, we would like to work with a model that is able to compensate the smallness of gravitational coupling. The simplest way is to consider only models with three-particle (two Galileons and one graviton) interaction, thus one may exclude the Paul term. One also should exclude this term as it experiences problems with the description of neutron stars \cite{Maselli:2016gxk}. And it is important to consider only three-particle interaction because of the classicalization \cite{Germani:2014hqa} (see \cite{Dvali:2016ovn} for a review). The idea of classicalization is the following: a non-renormalizable model may not require reformulation in terms of new degrees of freedom, but introduction of high-multiplicity interaction vertices, thus dissipating high-energy quanta via the large number of the same low-energy quanta \cite{Dvali:2010jz,Dvali:2011th,Dvali:2014ila}. Introduction of high-multiplicity interaction vertices may self-complete Galileons but at the same time prevent their manifestation in high-energy particle collision processes.

Summarizing all of the above, we conclude that only John term with a constant potential is worth to be considered as a leading Galileons contribution to GR:
\begin{equation}\label{John}
  S_{int}=\int d^4 x ~ \sqrt{-g} ~ \beta ~ G_{\mu\nu} ~ \nabla^\mu \phi ~\nabla^\nu\phi ~,
\end{equation}
where $\beta$ is coupling constant with dimension of inverse square mass. Thus $1/\sqrt{\beta}$ defines energy scale at which graviton-Galileons coupling becomes appreciable. As we had highlighted before, this energy scale must be smaller than Planck one but may not be bigger than electroweak scale. Galileons-graviton interaction are able to manifest themselves at low energies, but the manifestation will weakly affect Standard Model matter.

\begin{figure}[h]
\begin{center}
  \includegraphics[height=5cm]{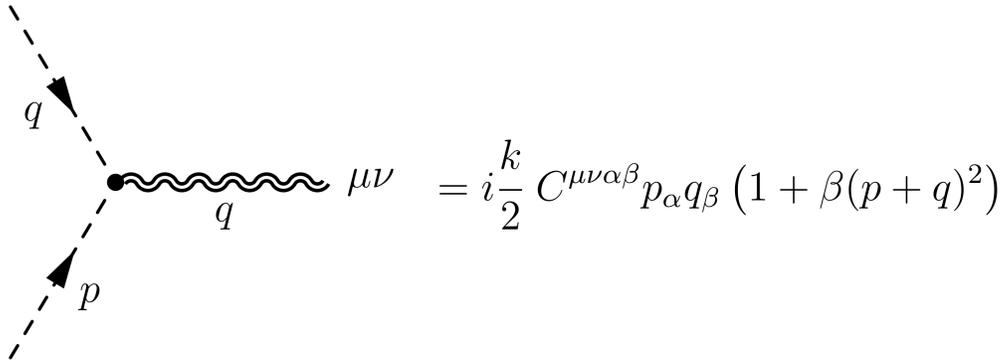}
\end{center}
\caption{Three-particle Galileons-graviton interaction}\label{Diagram}
\end{figure}

Following the standard procedure \cite{Veltman:1975vx,DeWitt:2007mi} we will treat gravity as small perturbations over flat background, Feynman rule for Galileons-graviton interaction vertex has the following form:
\begin{equation}
V^{\mu\nu} (p,q) =  i \frac{k}{2} ~ C^{\mu\nu\alpha\beta} p_\alpha q_\beta (1 + \beta (p+q)^2) ~,
\end{equation}
where $C^{\mu\nu\alpha\beta} = \eta^{\mu\alpha}\eta^{\nu\beta} + \eta^{\mu\beta}\eta^{\nu\alpha} - \eta^{\mu\nu}\eta^{\alpha\beta}$, $p$ and $q$ are scalar particles momentums and $k=\sqrt{8 \pi G}$ is the gravitational coupling. The corresponding diagram is placed in \Fref{Diagram}. Feynman rules for gravitational interaction with Standard Model matter may be found in \cite{Han:1998sg}.

As we had pointed out before, one must stick to tree-level diagrams in order to stand within the area of applicability of Galileons. We also want to find the manifestation of Galileons in collision experiments. This leaves a small room for processes worth researching, namely, this are annihilation processes of either fermions or vector particles to Galileons pairs. It is sensible to use LHC data to search for the processes in neutron-neutron scattering, such processes may be found by a loss of transfered momentum as our detectors are not able to detect Galileons. Thus we need to calculate the amplitude of fermion--anti-fermion and vector--anti-vector annihilation to Galileons.

\begin{figure}[h]
  \begin{center}
    \includegraphics[height=5cm]{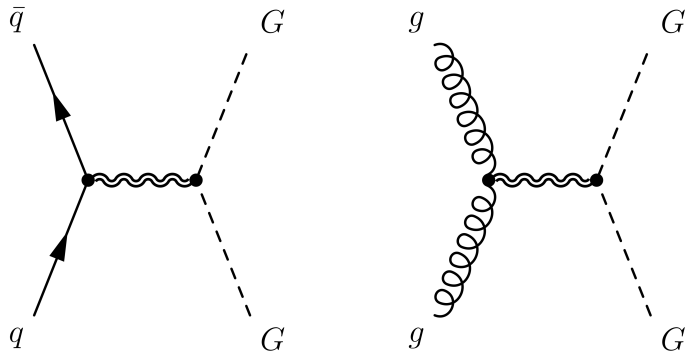}
  \end{center}
\caption{Digrams of lepton and quark annihilation to two Galileons. $q$ --quark, $\bar q$ -- anti-quark, $g$ -- gluon, $G$ -- Galileon.}
\end{figure}

As we are working within high-energy limit we will treat all particles as massless. Then, the cross-section of fermion--anti-fermion annihilation takes the following form:
\begin{eqnarray}\label{s_ff_0}
  \sigma_{f\bar f} = \frac{E^2 k^4}{15360 \pi} ~\left(1+\beta E^2\right)^2 ~,
\end{eqnarray}
where $E$ is center-of-mass energy. The cross-section of vector--anti-vector annihilation to Galileons has the form:
\begin{equation}\label{s_vv_0}
  \sigma_{v \bar v} = \frac{7 E^2 k^4}{15360 \pi} \left( 1 + \beta E^2\right)^2  ~.
\end{equation}
It would be convenient to use different notations via introduction of Galileons energy $E_G$ defined by the following:
\begin{equation}
  E_G = \frac{1}{\sqrt{| \beta |}} ~,
\end{equation}
and we are going to call $E_G$ Galileon energy hereafter. In this notations it is possible to rewrite expressions \eref{s_ff_0} and \eref{s_vv_0} in the following form:
\begin{eqnarray}
  \sigma_{f \bar f} = \frac{\pi}{60} ~ \frac{1}{m_p^2} ~  S\left(\frac{E}{E_G}\right) ~, \label{s_ff} \\
  \sigma_{v \bar v} = \frac{7\pi}{60} ~ \frac{1}{m_p^2} ~ S\left(\frac{E}{E_G}\right)  ~, \label{s_vv} \\
  S\left(\frac{E}{E_G}\right) = \left(\frac{E_G}{m_p}\right)^2 ~\left( \frac{E}{E_G} \right)^2 ~\left( 1 \pm \left(\frac{E}{E_G}\right)^2 \right)^2 \label{S_func} ~,
\end{eqnarray}
in expression \eref{S_func} plus sign corresponds to the positive value of Galileons coupling, minus -- to the negative value. We separated dimensionless modulation function $S(E/E_{G})$ \eref{S_func} for the sake of simplicity as it modulates both cross-sections. The sign of Galileons coupling affects behavior of cross-section only in the area of small $E/E_G$ and, as we will demonstrate, one is not able to distinguish the sign of $\beta$ at the current level of precision.

\begin{figure}[h]
  \begin{minipage}[h]{0.5\linewidth}
    \center{\includegraphics[width=8cm]{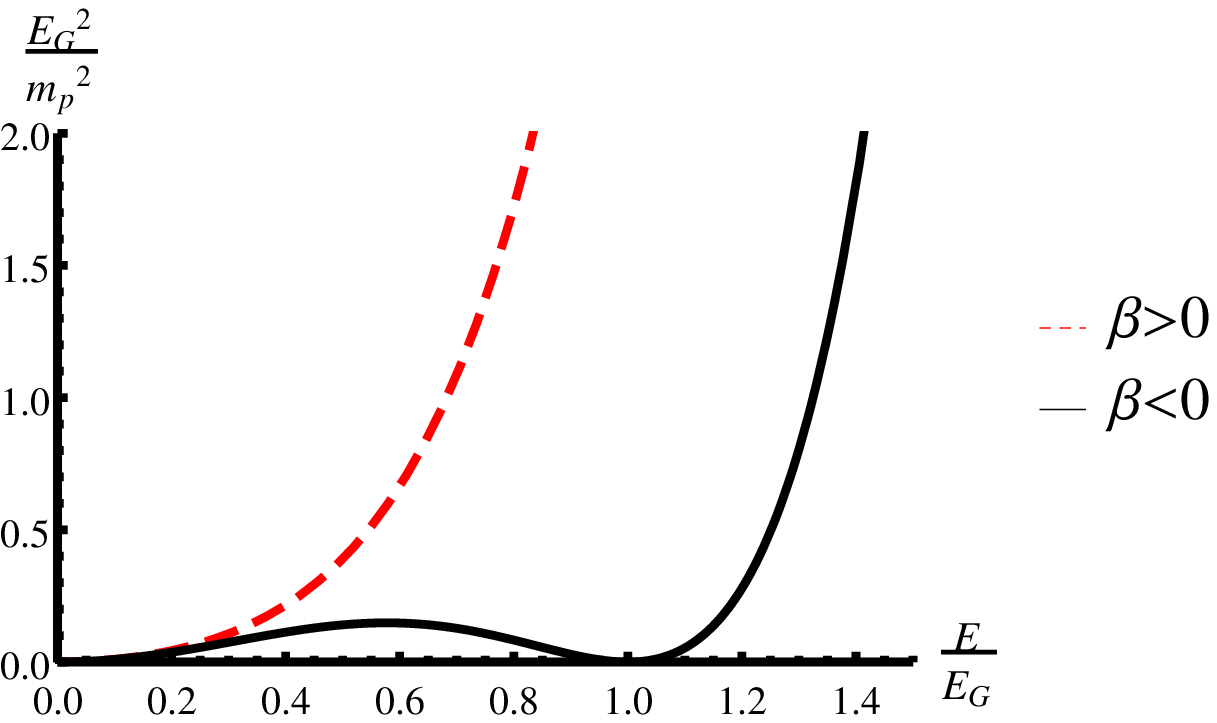}}
  \end{minipage}
  \begin{minipage}[h]{0.5\linewidth}
    \center{\includegraphics[width=9cm]{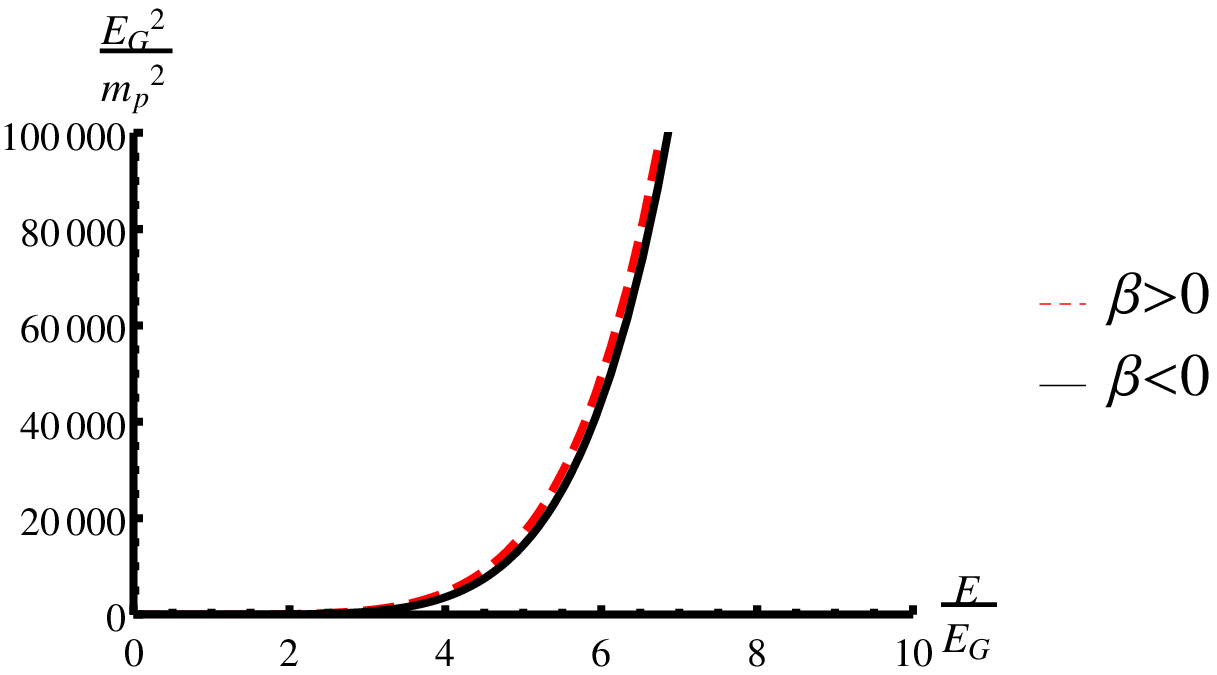}}
  \end{minipage}
  \caption{Modulation function \eref{S_func} in area of small and large $E/E_G$.}\label{Plot1}
\end{figure}

It is possible to establish elementary constraints on the value of $E_G$. Both cross-sections are suppressed by gravitational coupling so we would be able to constraint only large values of $E/E_G$. Second, we would like to use LHC data for the constraints, thus we need to consider neutron-neutron scattering processes. At high energies gluons dominate in neutron, so we would consider \eref{s_vv} as a leading order contribution to Galileon creation process. We also should multiply \eref{s_vv} by the factor of eight corresponding to the number of gluons. Thus, the leading contribution to Galileon creation processes in neutron-neutron scattering will have the following form:
\begin{eqnarray}\label{sigma}
  \sigma_{nn}=\frac{14}{15} \frac{1}{m_p^2} \left( \frac{E_G}{m_p} \right)^2 ~\left( \frac{E}{E_G} \right)^6 ~.
\end{eqnarray}
In order to constrain the value of the cross-section \eref{sigma} we take LHC data on Higgs creation \cite{Heinemeyer:2013tqa}. As a proper reference point we take a data on WH-production processes at CM energy $\sqrt{s}=8$ TeV, the smallest cross-section of the processes is $7.7$ fb. This results in the following constraints:
\begin{eqnarray}
  E_G > 10^{-19} {GeV} ~, \label{Constraint_1}\\
  \beta < 10^{36} {GeV}^{-2} = 10^8 {cm}^2 = 10^{-13} {g}^{-2} ~. \label{Constraint_2}
\end{eqnarray}

\section{Discussion and conclusion}

Summarizing the results, we would like to say the following. Galileons and especially the Fab Four are perspective modification of GR. They can provide description of both inflation and late-time acceleration \cite{Starobinsky:2016kua}, black holes \cite{Rinaldi:2012vy,Babichev:2013cya,Charmousis:2015aya,Babichev:2015rva,Anabalon:2013oea,Minamitsuji:2013ura} and stars \cite{Maselli:2016gxk,Brihaye:2016lin,Cisterna:2015yla,Cisterna:2016vdx}. It is important to search for beyond GR effects in Galileons and verify them. On the level of classical black hole-like solutions it is not always possible, as Galileons admit stealth Schwarzschild solution \cite{Babichev:2013cya}. At the level of classical cosmological solutions one may verify the primordial perturbation spectrum of Galileons \cite{Regan:2014hea,Arroja:2013dya}. In this article we discussed the high-energy manifestation of Galileons in collision experiments. We discussed the basic principles that we used to narrow the Fab Four class to one particular model. We considered only models with kinetic coupling that admit three-particle (two Galileons and one graviton) interaction. We excluded models with high-multiplicity Galileons interaction as classicalization may appear in those models. Certainly, classicalization may allow Galileons to form self-complete non-remormalizable model of Gravity in high-energy regime, but it also rules out any desirable manifestation of Galileons in collision experiments. Finally, we established elementary constraints on the model coupling using LHC data on Higgs production processes (WH-production to be precise). As Galileon production processes are suppressed by gravitational coupling, the constraints \eref{Constraint_1},\eref{Constraint_2} are pretty weak. The constraint \eref{Constraint_2} may be improved by using parton distribution function, but this will not change the general picture and new constraints will also be weak.

\ack
This work was supported by Russian Foundation for Basic Research via grant RFBR \textnumero 16-02-00682. The author is sincerely grateful to S.O. Alexeyev for useful discussion.
\vspace{1cm}

\bibliographystyle{unsrt}
\bibliography{PhysicsArticle}

\end{document}